# Electronic catalogue of muonic X-rays

*Daniya* Zinatulina[1,*], *Chantal* Briançon[2], *Victor* Brudanin[1], *Viacheslav* Egorov[1], *Lev* Perevoshchikov[1], *Mark* Shirchenko[1], *Igor* Yutlandov[1], and *Claude* Petitjean[3]

[1]Joint Institute for Nuclear Research, DLNP, 141980 Dubna, Russia
[2]Centre de Spectrometrie Nucleaire et de Spectrometrie de Masse, 91405 Orsay, France
[3]Paul Sherrer Institute, CH 5232 Villingen, Switzerland

**Abstract.** $\mu$X-ray spectra for Z=9-90 were measured with HPGe detectors and muonic beams of PSI (Villigen, Switzerland) [1]. The results are presented as electronic atlas composed of graphic plots. The atlas is available at JINR site [2].

## 1 Experiment and purposes

Stopping of negative muons in mater is accompanied by emission of so-called "muonic X-rays". This radiation, similar to conventional electronic characteristic X-rays, is unique for each chemical element, but has much higher energy (up to ~10 MeV). Contrary to electronic X-rays, there was lack of systematic information about $\mu$X-rays. We have tried to fill this gap.

Idea of the experiment is based on the precise measurement of time- and energy-distribution of γ-rays following μ-capture. Low-momentum negative muons from the μE4 or μE1 beam-line of the PSI "meson factory" were stopped in thin targets (about 0.5 g/cm$^2$). In case of gas target, we used a special construction, described in our previous works [3, 4].

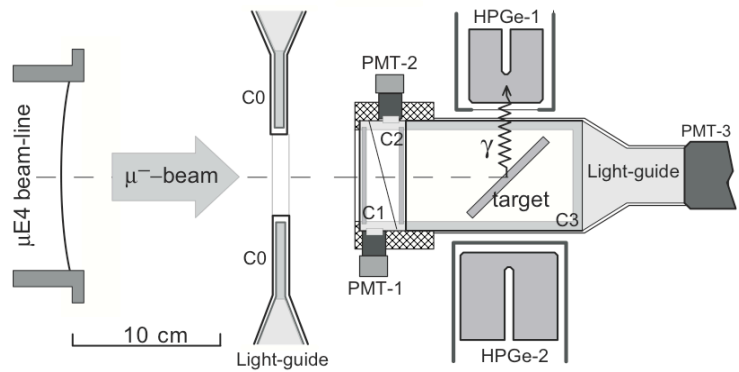

**Fig. 1.** Measurement set-up.

---

[*] Corresponding author: d.zinatulina@gmail.com

Muonic X-rays accompanying muon stops were registered with several HPGe detectors (description of the acquisition system was given in [3]). The experimental set-up is shown in Figure 1.

There are ~75 elements presented in our catalogue. We always tried to use the simple chemical compositions for the investigation. In case when the element was too chemically active, we used it as a passive compound. The second "companion" was chosen to have Z which differs a lot from the main one, in order to avoid the overlapping of corresponding muonic spectra.

Below there is a Figure 2 with the periodic table (from the web-page with our catalogue), where the different colors of the elements stand for the different type of the target: "green" – target is comprised from the pure element, "red" – target is used as an oxide, "orange" – halogen compound, "violet" – nitrate, "grey" – not investigated. To see the spectra of the interesting element, one need to click on it in Periodic table.

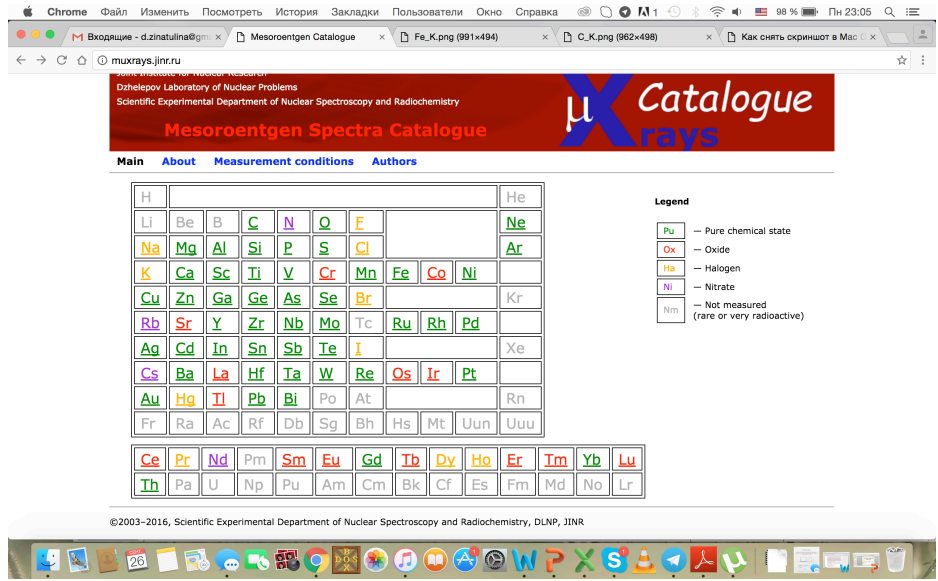

**Fig.2** Periodic table from the web-page of the http://muxrays.jinr.ru

## 2 Results

As a result of the measurements, we have what we call "prompt energy spectra" ($T_{\mu\text{-}\gamma}$< 20-50 ns) which contain mainly muonic X-rays, fast part of delayed gammas and scattered beam electrons. Normally, muon can be captured by nucleus from the upper orbits only in very heavy muonic atoms, so for the light and intermediate Z it could be stated that at-least 99% of μ-stops are accompanied by any of *np-1s* muonic transition. Our interactive catalogue contains the spectra of *np-1s* muonic transition for all of the measured elements. For many of the elements we also present L-, M-series and sometimes even more (N, O etc). As example, such spectra are presented in the Figures 3, 4. Figure 3 is a total μX-ray spectrum of Cd. The structure of the spectrum sometimes quite complicated. There are the Compton bump, pair production, and also SE (single escape) and DE (double escape) peaks due to high energy of the resulting gamma. Figure 4 – is a zoom of the K-series from the total µX-ray spectrum in Figure 3 (that could be done as well for other L-, M- etc. series also by clicking them in Total spectrum).

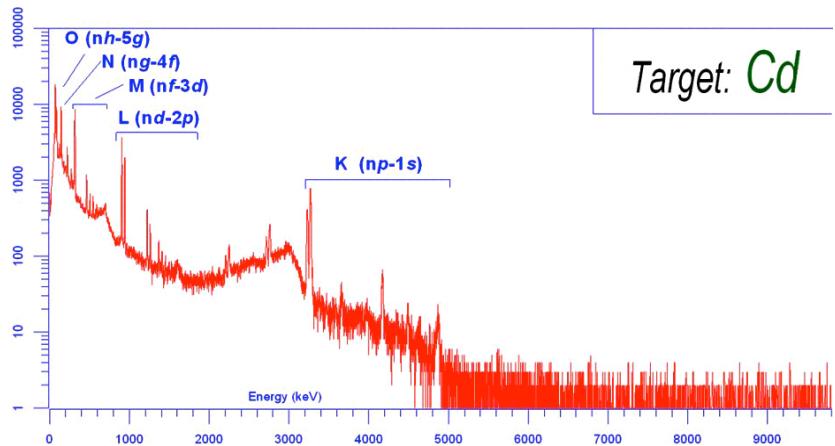

**Fig. 3.** Total µX-ray spectrum for Cd.

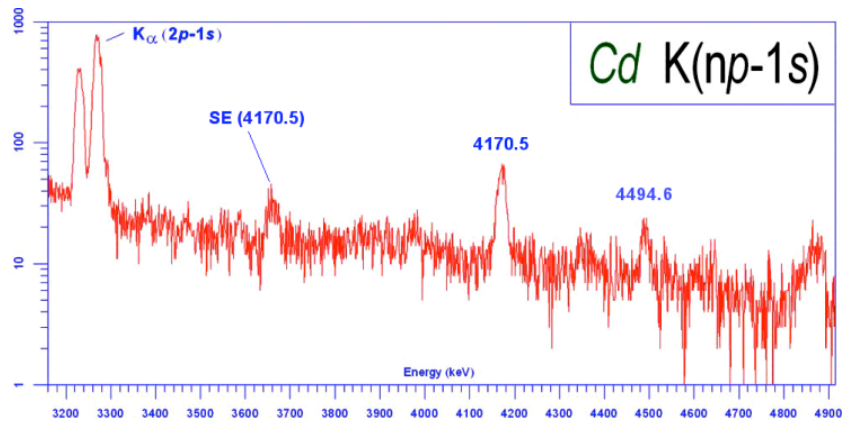

**Fig. 4.** K-series of Cd.

Whereas for muons the radius of K-shell comparable of the nuclei radius, then this K-shell very sensitive to the nucleus parameters. That is why µX-transition to K-shell depend on the properties of the nucleus. Thus, for K-series of the Cd, for example, we have a complex structure of the Kα, because there is an isotopic shift. The region of such Kα is presented in Figure 5.

## 3 Conclusions

Finally, we combined all the information into the catalogue and put it on the web-page [2]. In the Figure 6 the main web-page of the Electronic catalogue is shown. There is the information about the conditions of the measurements, contacts of the authors, users guide etc. The information from the µX-ray spectra catalogue is important for the precise identification of the structure of the background and for the correct selection of the targets and construction materials for different experiments with muons. Total intensity of the

μX(Z)-ray K-series also give the number of muons stopped exactly in the target with specific Z.

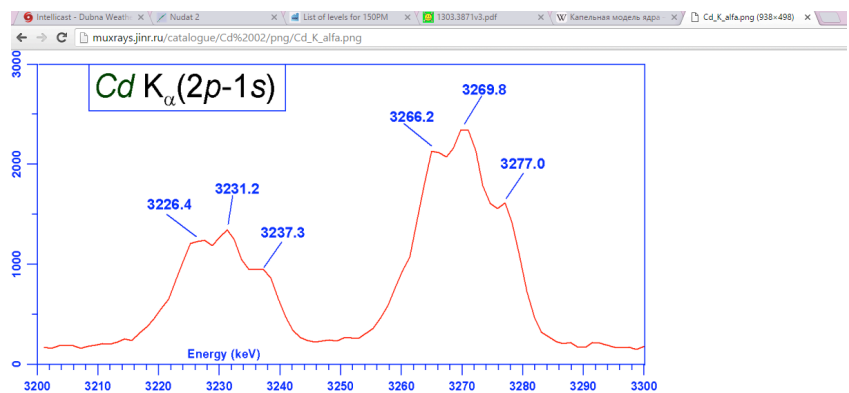

**Fig. 5.** Kα of Cd.

The enrichment of the measured isotopes could be found knowing the intensities of the Kα from the isotopic shifts. The information from that catalogue could be helpful for people who connected with mesoroentgen measurements.

**Fig. 6.** Main web-page of the http://muxrays.jinr.ru